# Energy-Efficient Power Control in Multipath CDMA Channels via Large System Analysis

Stefano Buzzi, *IEEE, Senior Member*, Valeria Massaro, *IEEE, Student Member*, and H. Vincent Poor, *IEEE, Fellow*

***Abstract*— This paper is focused on the design and analysis of power control procedures for the uplink of multipath code-division-multiple-access (CDMA) channels based on the large system analysis (LSA). Using the tools of LSA, a new decentralized power control algorithm aimed at energy efficiency maximization and requiring very little prior information on the interference background is proposed; moreover, it is also shown that LSA can be used to predict with good accuracy the performance and operational conditions of a large network operating at the equilibrium over a multipath channel, i.e. the power, signal-to-interference-plus-noise ratio (SINR) and utility profiles across users, wherein the utility is defined as the number of bits reliably delivered to the receiver for each energy-unit used for transmission. Additionally, an LSA-based performance comparison among linear receivers is carried out in terms of achieved energy efficiency at the equilibrium. Finally, the problem of the choice of the utility-maximizing training length is also considered. Numerical results show a very satisfactory agreement of the theoretical analysis with simulation results obtained with reference to systems with finite (and not so large) numbers of users.**

## I. Introduction

A recent trend in the design of wireless networks is to use game theoretic tools to model non-cooperative behavior between network entities competing for a common resource and aiming at maximizing an assigned utility measure [1], [2]. As an example, studies [3], [4] show how game theory can be used in the uplink of a code division multiple access (CDMA) wireless data network to develop non-cooperative resource allocation algorithms, and, *in primis*, power control procedures, aimed at energy efficiency maximization, defined as the ratio of the number of information symbols successfully delivered to the receiver for each Joule of energy used for transmission. The approach therein pursued has been recently extended in [5] to the relevant case of an asynchronous, bandlimited multiuser system operating over multipath fading channels. Implementation of the resulting resource allocation algorithms, however, although being non-cooperative, requires knowledge of the uplink signal-to-interference plus noise ratio (SINR) for each user, which, in turn, must be either estimated or computed based on some prior knowledge about the interference background.

In the paper [6], instead, some of the authors have shown that large system analysis (LSA) [7], [8] may be used to obtain simple power control procedures that require knowledge only of the channel gain for the user of interest, with no need for the interference parameters. In particular, it is also shown that LSA may be used to predict the operational conditions of a large network[1], i.e. the transmit power profile, the achieved utility profile, and the achieved SINR profile across users. The above results, however, refer to the idealized scenario of a synchronous system with frequency-flat fading.

In this paper, instead, we enhance the approach of [6] to deal with the more realistic case of an *asynchronous*, *bandlimited* wireless CDMA network operating over a *frequency-selective* fading channel. The contribution of this paper can be summarized as follows.

- We show how LSA permits the development of simple utility-maximizing power control algorithms amenable to a decentralized implementation and requiring knowledge of the channel coefficients for the user of interest only; we consider also the case of constrained maximum transmit power for each mobile terminal.
- We use LSA to derive simple closed-form expressions for the asymptotic utility and transmit power profile across users for a system with a large number of users.
- We carry out an LSA-based performance comparison, in terms of energy-efficiency at the equilibrium, between the classical linear receivers.
- The interesting issue of the optimal (in the sense of utility-maximizing) choice of the training length for enabling estimation of the channel coefficients is also considered.

Numerical results will show that the proposed procedures are effective and that the LSA-based performance results hold with satisfactory accuracy also when the network is not particularly large.

## II. Preliminaries and Problem Statement

Consider the uplink of an asynchronous direct-sequence (DS)-CDMA system with $K$ users, employing bandlimited square-root raised cosine chip pulses and operating over a frequency-selective fading channel. The received signal at the access point (AP) may be written as[2]

$$r(t) = \sum_{p=0}^{B-1} \sum_{k=1}^{K} \sqrt{p_k} b_k(p) s'_k(t-\tau_k - pT_b) * c_k(t) + w(t) \, . \quad (1)$$

S. Buzzi and V. Massaro are with DAEIMI, University of Cassino, Italy; e-mail: {buzzi, massaro}@unicas.it. H. V. Poor is with the School of Engineering and Applied Science, Princeton University, NJ, USA; e-mail: poor@princeton.edu. The research was supported in part by the U. S. National Science Foundation under Grants ANI-03-38807 and CNS-06-25637.

[1]The term *large* refers to the case in which both the number of users and the system processing gain grow large with their ratio fixed.

[2]For the sake of simplicity we assume here a real signal model; however, the extension to complex signals to account for I and Q components is straightforward.

In the above expression, $B$ is the transmitted frame or packet length, $T_b$ is the bit-interval duration, $p_k$ and $\tau_k \geq 0$ denote the transmit power and timing offset of the $k$-th user, and $b_k(p) \in \{+1, -1\}$ is the $k$-th user's information symbol in the $p$-th signaling interval (extension to modulations with a larger cardinality is straightforward). Moreover, $c_k(t) = \sum_{l=0}^{L-1} \alpha_{k,l}\delta(t - \tau_{k,l})$ is the impulse response modeling the multipath channel between the receiver and the $k$-th user's transmitter, with $\alpha_{k,l}$ and $\tau_{k,l}$ representing the channel gain and the timing offset associated with the $l$-th replica from the $k$-th user's transmitter, respectively, and $L$ the number of multipath replicas. Moreover, $w(t)$ is the additive noise term, which is assumed to be a zero-mean, white Gaussian process with Power Spectral Density (PSD) $\mathcal{N}_0/2$, and $s'_k(t)$ is the $k$-th user's signature waveform. The receiver front-end consists of a filter matched to the square-root raised cosine waveform, followed by a chip-rate sampler. Assume now that the samples at the output of the receiver front-end and corresponding to the interval $\mathcal{I}_p^k = [\tau_k + \min_l \tau_{k,l} + pT_b, \tau_k + \min_l \tau_{k,l} + (p+1)T_b]$ are used to detect the data symbol $b_k(p)$; stacking such samples in the vector $\boldsymbol{y}_k(p)$, it is easy to show that[3]

$$\begin{aligned}\boldsymbol{y}_k(p) = & \sum_{l=0}^{L-1} \sqrt{p_k} \left[\alpha_{k,l} b_k(p-1)\boldsymbol{s}_{k,l}^{(-1)} + \alpha_{k,l} b_k(p)\boldsymbol{s}_{k,l}\right] + \\ & \sum_{h \neq k} \boldsymbol{z}_{h,k}(p) + \boldsymbol{n}_k(p) \ .\end{aligned} \tag{2}$$

In the above expression, $\boldsymbol{s}_{k,l}^{(-1)}$ and $\boldsymbol{s}_{k,l}$ are the discrete-time versions of the $l$-th signal replica associated to the information symbols $b_k(p-1)$ and $b_k(p)$, respectively, and windowed to the interval $\mathcal{I}_p^k$; the vectors $\boldsymbol{z}_{h,k}(p)$ and $\boldsymbol{n}_k(p)$ are the $h$-th user's interference and the thermal noise contributions windowed to $\mathcal{I}_p^k$, respectively.

Assume now that each mobile terminal is interested both in having its data received with as small as possible error probability at the AP, and in making optimal use of the energy stored in its battery. Obviously, these are conflicting goals, since error-free reception may be achieved by increasing the received SNR, i.e. by increasing the transmit power, which of course comes at the expense of battery life[4]. Following discussion in [3], [4], [5], a useful approach to quantify these conflicting goals is to define the utility $u_k$ of the $k$-th user as

$$u_k = R\frac{B - N_T}{B}\frac{f(\gamma_k)}{p_k} \ , \quad \forall k = 1, \ldots, K \ , \tag{3}$$

wherein $R$ is the common data-rate of the network, $N_T$ is the number of overhead bits, reserved, e.g., for channel estimation and/or parity checks, $\gamma_k$ is the $k$-th user's SINR, and the *efficiency function* $f(\gamma_k) = (1 - e^{-\gamma_k/2})^B$. Note that the utility definition (3) is an analytically tractable approximation of the number of bits successfully delivered to the receiver for each Joule of energy used for transmission; as a consequence, maximization of (3) is approximately equivalent to maximization of the energy-efficiency of the transmitter.

A decision on the information symbol $b_k(p)$ is taken at the AP according to the rule

$$\widehat{b}_k(p) = \text{sign}\left[\boldsymbol{x}_k^T \boldsymbol{O}_k^T \boldsymbol{y}_k(p)\right] \ , \tag{4}$$

wherein $\boldsymbol{O}_k$ is an $N \times (N-1)$-dimensional orthonormal matrix whose column span is orthogonal to the intersymbol interference (ISI) vector $\sum_{l=0}^{L-1} \alpha_{k,l}\boldsymbol{s}_{k,l}^{(-1)}$, and $\boldsymbol{x}_k$ is an $(N-1)$-dimensional vector to be suitably designed. We can thus consider a non-cooperative game wherein each user selfishly tries to maximize its own utility by varying its own transmit power $p_k$ (subject to the constraint $p_k \leq P_{k,\max}$) and its uplink receiver $\boldsymbol{x}_k$, i.e.

$$\max_{p_k, \boldsymbol{x}_k} \frac{f(\gamma_k(p_k, \boldsymbol{x}_k))}{p_k} = \max_{p_k} \frac{f\left(\max\limits_{\boldsymbol{x}_k} \gamma_k(p_k, \boldsymbol{x}_k)\right)}{p_k} \ . \tag{5}$$

Following the approach of [5], the following result can be proven.

**Proposition:** *Let $\boldsymbol{M}_{\boldsymbol{y}_k\boldsymbol{y}_k}$ denote the covariance matrix of the vector $\boldsymbol{y}_k(p)$. Let $\boldsymbol{h}_{k,0} = \sum_\ell \alpha_{k,\ell}\boldsymbol{s}_{k,\ell}$. The non-cooperative game defined in (5) admits a unique Nash equilibrium point $(p_k^*, \boldsymbol{x}_k^*)$, for $k = 1, \ldots, K$, wherein*

- *$\boldsymbol{x}_k^* = \sqrt{p_k}\left(\boldsymbol{O}_k^T \boldsymbol{M}_{\boldsymbol{y}_k\boldsymbol{y}_k}\boldsymbol{O}_k\right)^{-1}\boldsymbol{O}_k^T\boldsymbol{h}_{k,0}$ is the unique (up to a positive scaling factor) $k$-th user's receive filter that maximizes its SINR $\gamma_k$. Denote $\gamma_k^* = \max_{\boldsymbol{x}_k}\gamma_k$.*
- *$p_k^* = \min\{\bar{p}_k, P_{k,\max}\}$, with $\bar{p}_k$ the $k$-th user's transmit power such that the $k$-th user's maximal SINR $\gamma_k^*$ equals $\bar{\gamma}$, i.e. the unique solution of the equation $f(\gamma) = \gamma f'(\gamma)$, with $f'(\gamma)$ denoting the derivative of $f(\gamma)$.*

**Proof:** The proof is omitted due to lack of space. ■

The above equilibrium can be reached according to the following procedure: *(a)* for a given set of users' transmit powers, the receiver filter coefficients can be set according to the relation $\boldsymbol{x}_k^* = \sqrt{p_k}\left(\boldsymbol{O}_k^T \boldsymbol{M}_{\boldsymbol{y}_k\boldsymbol{y}_k}\boldsymbol{O}_k\right)^{-1}\boldsymbol{O}_k^T\boldsymbol{h}_{k,0}$; and *(b)* each user can then tune its power so as to achieve the target SINR $\bar{\gamma}$. These steps are repeated until convergence is reached.

## III. LSA-BASED POWER CONTROL

As already discussed, implementation of the above task *(b)* requires knowledge of the uplink SINR for each user, which in turn, must be either estimated or computed based on prior information about the current status of the multiuser interference. In order to overcome this limitation, in the following we propose an LSA-based utility maximizing power control algorithm. We assume here that each user knows its own channel coefficients and that there is no constraint on the maximum transmit power; both hypotheses will be removed later on in the paper.

First of all, note that according to [8] in a CDMA network with negligible ISI wherein both the number of users $K$ and the processing gain $N$ grow large but with their ratio $K/N =$

[3]For the sake of simplicity, we assume here that the multipath delay spread is such that intersymbol interference from only one previous symbol affects data detection for each user. This assumption is not crucial and can be removed straightforwardly.

[4]Of course there are many other strategies to lower the data error probability, such as for example the use of error correcting codes, diversity exploitation, and implementation of optimal reception techniques at the receiver. Here, however, we are mainly interested in energy efficient data transmission and power usage, so we assume that only the transmit power and the receiver strategy can be varied to achieve energy efficiency.

$\alpha$ fixed, and with randomly chosen unit-energy signature waveforms, the $k$-th user's SINR at the output of a minimum mean-square error (MMSE) linear receiver converges almost surely (a.s.) to a deterministic quantity $\gamma_k^*$, which is expressed as

$$\gamma_k^* = \frac{\sum_{l=1}^{L} p_k |\hat{\alpha}_{k,l}|^2 \beta_d}{1+\xi_k^2 \beta_d}, \tag{6}$$

wherein $\hat{\alpha}_{k,l}$ is the estimate of the channel gain $\alpha_{k,l}$, $\xi_k^2$ is the channel estimation error variance for the $k$-th user[5], and $\beta_d$ is the unique solution of the equation

$$\beta_d = \left[\frac{\mathcal{N}_0}{2} + \frac{1}{N}\sum_{h\neq k, h=1}^{K} \left((L-1)I(\xi_h^2, \beta_d) + I\left(\sum_{l=1}^{L} p_h |\alpha_{h,l}|^2 + \xi_h^2, \beta_d\right)\right)\right]^{-1}, \tag{7}$$

with $I(a,b) = a/(1+ab)$.

Now, since there is no constraint on the maximum transmit power, it is reasonable to assume that all the users are able to achieve the target SINR $\bar{\gamma}$ and that all the users are to be received with the same power $P_R$, i.e.

$$p_1 \sum_{l=0}^{L-1} |\alpha_{1,l}|^2 = \ldots = p_K \sum_{l=0}^{L-1} |\alpha_{K,l}|^2 = P_R \, . \tag{8}$$

On substituting Eqs. (8) and (7) into Eq. (6), it is easy to show that $P_R = \frac{\bar{\gamma}\mathcal{N}_0/2}{(1-\alpha\frac{\bar{\gamma}}{1+\bar{\gamma}})}$, thus implying that

$$p_k = \frac{\bar{\gamma}\mathcal{N}_0/2}{\left(\sum_{l=0}^{L-1} |\alpha_{k,l}|^2\right)(1-\alpha\frac{\bar{\gamma}}{1+\bar{\gamma}})} \, , \quad k=1,\ldots,K \, , \tag{9}$$

wherein the feasibility condition $\alpha < \frac{1+\bar{\gamma}}{\bar{\gamma}}$ must hold. Equation (9) represents the desired decentralized power control rule; note that each user may implement this rule based on the knowledge of only its own channel coefficients.

LSA can also be used to predict the utility profile achieved in a large network. Indeed, substituting Eq. (9) into (3) we have

$$u_k = R\frac{B-N_T}{B}\frac{f(\bar{\gamma})\left(\sum_{l=0}^{L-1} |\alpha_{k,l}|^2\right)(1-\alpha\frac{\bar{\gamma}}{1+\bar{\gamma}})}{\bar{\gamma}\mathcal{N}_0/2} \, . \tag{10}$$

The above equation represents an LSA-based approximation for the energy-efficiency for the $k$-th user. Again, it can be computed based on the knowledge of the channel coefficients for the user of interest. A further step is then the computation of the utility profile with no prior knowledge about the channel coefficient realizations of the system. To this aim, recall that in [9] the following result has been stated.

**Lemma**: *Let $x_1, \ldots, x_K$ be a sequence of independent and identically distributed (i.i.d) random variates with cumulative distribution function (CDF) $F(\cdot)$, and let $x_{[1]}, x_{[2]}, \ldots, x_{[K]}$ be the same random variates sorted in non-increasing order. Then we have that $x_{[\ell]}$ converges, for increasing $K$, in probability to $F^{-1}(\frac{K-\ell}{K})$, $\forall l = 1, \ldots, K$.*

[5]Since we are assuming that the channel coefficients are perfectly known, we have here $\xi_k^2 = 0$.

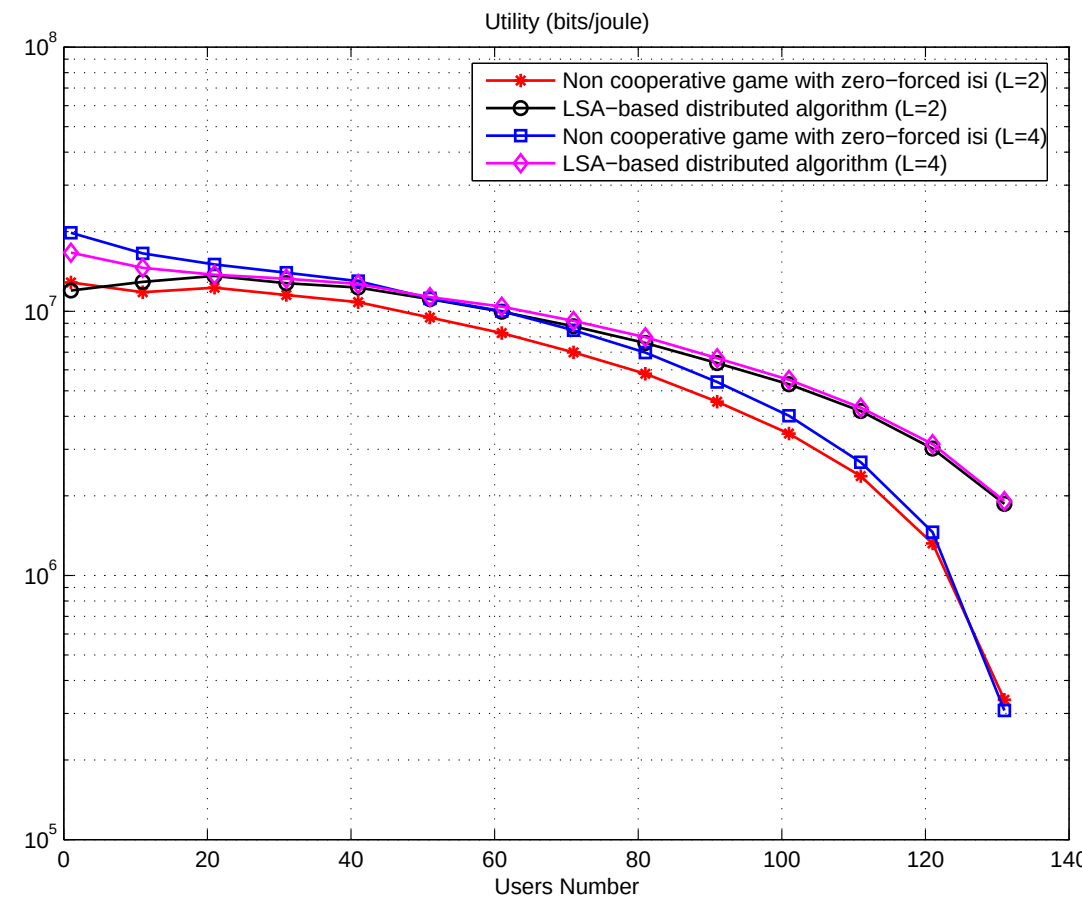

Fig. 1. Achieved average utility versus number of active users for the LSA-based proposed algorithm and for the non-cooperative game with zero-forced ISI. Two different values of $L$ have been considered.

The above lemma asserts that if we sort a large number of identically distributed random variates, we obtain a vector that is approximately equal to the uniformly sampled version of the inverse of the common CDF of the random variates. Applying this lemma to the channel coefficients $\sum_{l=0}^{L-1} |\alpha_{k,l}|^2$ we are thus able to obtain a rough estimate of the channel coefficient realizations in the network based only on their first-order statistics, and, thus, to predict the utility and power profile in the network according to the formulas[6]

$$p_k = \frac{\bar{\gamma}\mathcal{N}_0/2}{F^{-1}((K-k)/K)\left(1-\alpha\frac{\bar{\gamma}}{1+\bar{\gamma}}\right)} \, , \quad k=1,\ldots,K \, , \tag{11}$$

and

$$u_k = R\frac{B-N_T}{B}\frac{f(\bar{\gamma})F^{-1}((K-k)/K)(1-\alpha\frac{\bar{\gamma}}{1+\bar{\gamma}})}{\bar{\gamma}\mathcal{N}_0/2} \, , \tag{12}$$

for $k = 1, \ldots, K$ and with $F(\cdot)$ the CDF of $\sum_{l=0}^{L-1} |\alpha_{k,l}|^2$. Figure 1 shows the average utility versus the number of users for the non-cooperative game detailed in Proposition 1 and for the outlined distributed power control procedure. A system with processing gain $N = 128$ has been taken, and two different values of $L$ have been considered. Figure 2, instead, shows the power profile and the achieved utility profile for a system with $K = 120$ users. It is seen that the LSA-based approximations are very tight.

## IV. Comparison of Linear Multiuser Detectors

The tools of LSA can also be used to perform a comparison - in terms of achieved energy-efficiency at the Nash equilibrium - between different linear receivers. This analysis may turn out to be useful in order to compare the performance of the linear multiuser receivers not only in terms of error probability and/or near-far resistance, as usually happens, but also in terms of energy-efficiency. In particular, we focus here on the matched filter, on the linear MMSE receiver and on the multipath-decorrelating receiver. Following [8], it can be shown that, in the large system limit, the SINR at the output of a matched filter converges a.s. to the deterministic quantity $\gamma_k^*$ specified in Eq. (6), with $\xi_k^2 = 0$ (perfect channel state information is assumed here) and with $\beta_d$ expressed as

$$\beta_d = \left[ \frac{\mathcal{N}_0}{2} + \frac{1}{N} \sum_{h=1, h\neq k}^{K} \sum_{l=1}^{L} p_h |\alpha_{h,l}|^2 \right]^{-1} . \quad (13)$$

[6]Of course we are able to predict the ensemble of the users' achieved utilities and transmit powers, but are not able to predict the utility achieved by a certain user, unless its channel coefficients are available.

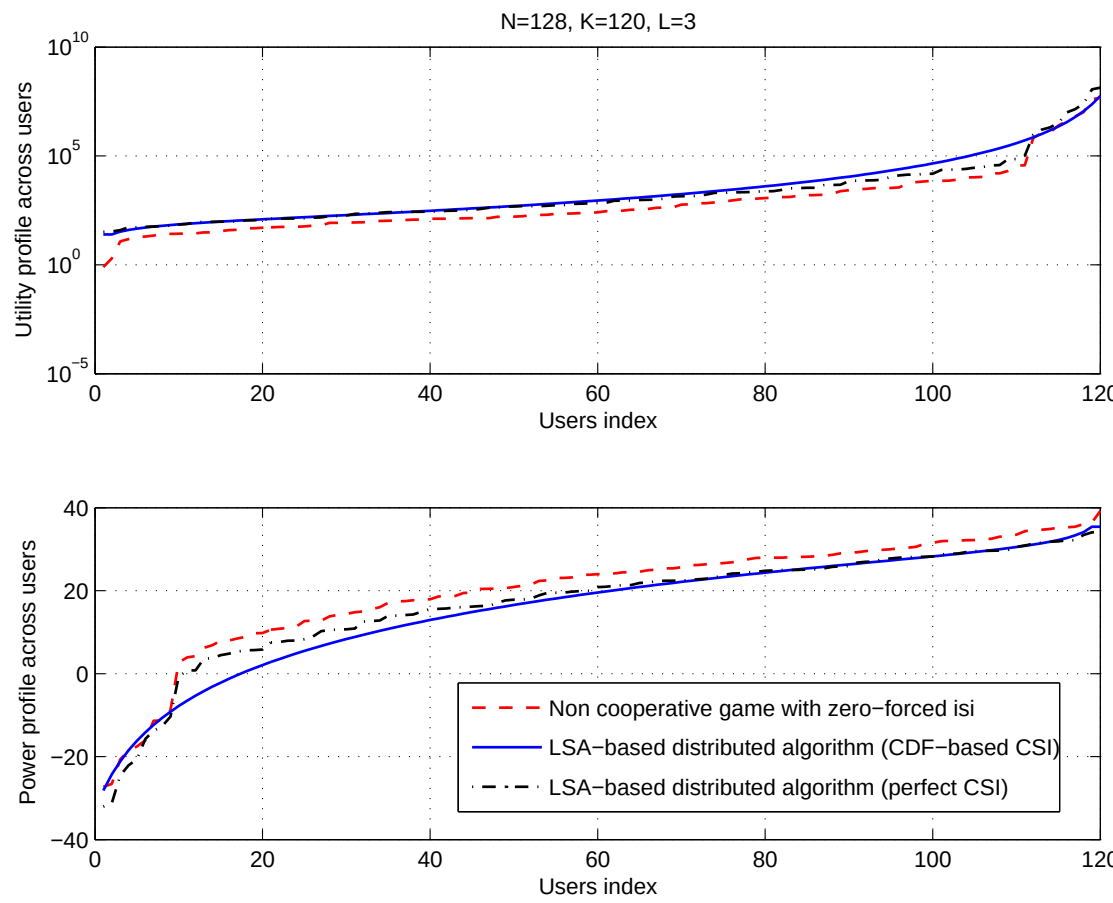

Fig. 2. Utility and power profiles across users as predicted by the LSA-based approximations and as obtained by direct implementation of the non-cooperative game with zero-forced ISI.

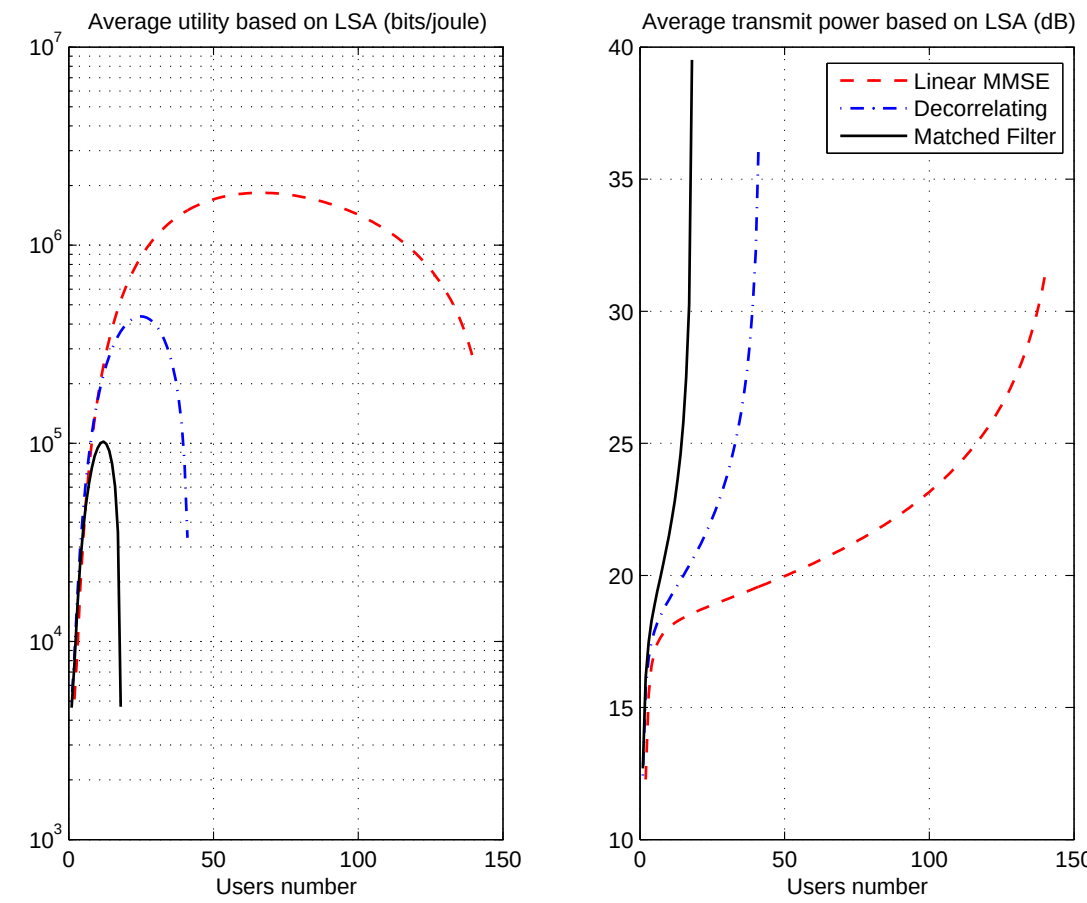

Fig. 3. Achieved average utility and average transmit power versus number of active users for three different linear receivers.

Substituting Eqs. (13) and (8) into Eq. (6), we obtain an expression for the $k$-th user's transmit power that is needed to achieve the target SINR, i.e.:

$$p_k^{\text{MF}} = \frac{\bar{\gamma}\mathcal{N}_0/2}{\sum_{l=0}^{L-1} |\alpha_{k,l}|^2 (1-\alpha\bar{\gamma})} , \quad (14)$$

wherein the feasibility condition $\alpha < \frac{1}{\bar{\gamma}}$ is to be fulfilled.

Consider now a decorrelating multipath-combining receiver; the asymptotic SINR is again given by Eq. (6) with $\xi_k^2 = 0$ and $\beta_d = \frac{1-\alpha L}{\mathcal{N}_0/2}$. Accordingly, we have

$$p_k^{\text{DEC}} = \frac{\bar{\gamma}\mathcal{N}_0/2}{\sum_{l=0}^{L-1} |\alpha_{k,l}|^2 (1-\alpha L)}, \quad (15)$$

with the feasibility condition $\alpha < \frac{1}{L}$. Now, using the lemma of the previous section we can again obtain the utility profile for a large CDMA network with matched filter or decorrelating multipath-combining receivers. We have thus

$$\begin{cases} u_k^{\text{MF}} = R\frac{B-N_T}{B}\frac{f(\bar{\gamma})F^{-1}((K-k)/K)}{\bar{\gamma}\mathcal{N}_0/2}\left(1-\alpha\bar{\gamma}\right), & \alpha < \frac{1}{\bar{\gamma}}; \\ u_k^{\text{DEC}} = R\frac{B-N_T}{B}\frac{f(\bar{\gamma})F^{-1}((K-k)/K)}{\bar{\gamma}\mathcal{N}_0/2}\left(1-\alpha\right), & \alpha < \frac{1}{L}. \end{cases} \quad (16)$$

In Fig. 3 we report the achieved average utility and the average transmit power versus the number of active users for the three considered linear receivers. The plot refers to a system with $L = 3$ paths and with processing gain $N = 128$. As expected, it is seen that the linear MMSE receiver achieves the largest energy-efficiency with the smallest average transmit power, and is capable of supporting far more users than the matched filter and the decorrelating receiver.

## V. Optimal training length for channel estimation

Let us now consider the practically relevant case in which the channel is estimated based on the $N_T$ training symbols. For MMSE channel estimation, and assuming that the average (with respect to the fading statistics) received power for each user equals $P$, the variance of the channel estimation error converges a.s., in the large system limit, to the quantity[7]

$$\xi^2 = \frac{P}{1+P\beta_c}, \quad (17)$$

where $\beta_c$ is the solution of

$$\beta_c = \left[ \frac{\mathcal{N}_0/2}{N_T} + \frac{\alpha L}{N_T}\frac{P}{1+P\beta_c} \right]^{-1} . \quad (18)$$

Substituting (17) into Eqs. (6) and (7) we are able to obtain the asymptotic SINR taking into account the channel estimation error. Of course, the larger $N_T$, the larger the achieved SINR, and, consequently, the user utility; on the other hand, since the energy efficiency also depends on the multiplicative term $(B-N_T)/B$, the utility of any user decreases to zero as $N_T \to B$. Accordingly, there is an optimal value of $N_T$ that maximizes the utility. The following procedure is thus aimed at evaluating utility as a function of $N_T$. First of all, we find the value of $\xi^2$ based on Eq. (17); this value is substituted into Eq. (7); moreover, we also let $p_h \sum_{l=0}^{L-1} |\alpha_{h,l}|^2 = P_R$ and, from Eq. (6), $P_R = \bar{\gamma}(1+\xi^2\beta_d)/\beta_d$. Putting together all these relations, we obtain a unique equation in $\beta_d$; let $\beta_d^*$ be the solution of this equation. Using again (6) we can thus obtain the power profile needed to achieve the target SINR $\bar{\gamma}$ when channel estimation errors are taken into account, i.e.

$$p_k = \bar{\gamma}\frac{1+\xi^2\beta_d^*}{\beta_d^* F^{-1}((K-k)/K)} , \quad k = 1, \dots, K . \quad (19)$$

Substituting the above relation into Eq. (3) provides the set of the achieved utilities for a given number of users $K$ and for assigned $N_T$, whose average then yields the achieved average utility. Repeating this procedure for several values of $N_T$ we can obtain the average utility versus $N_T$. The results of this procedure are reported in Fig. 4 for a system with $B = 120$ and for different values of $K$. Results confirm that there is an utility-maximizing $N_T$, which is represented in the plot by a circle.

[7]The assumptions that have been made imply that the channel estimation error variance is the same for all the users; relaxation of this hypothesis can be done straightforwardly, and is not presented here due to lack of space.

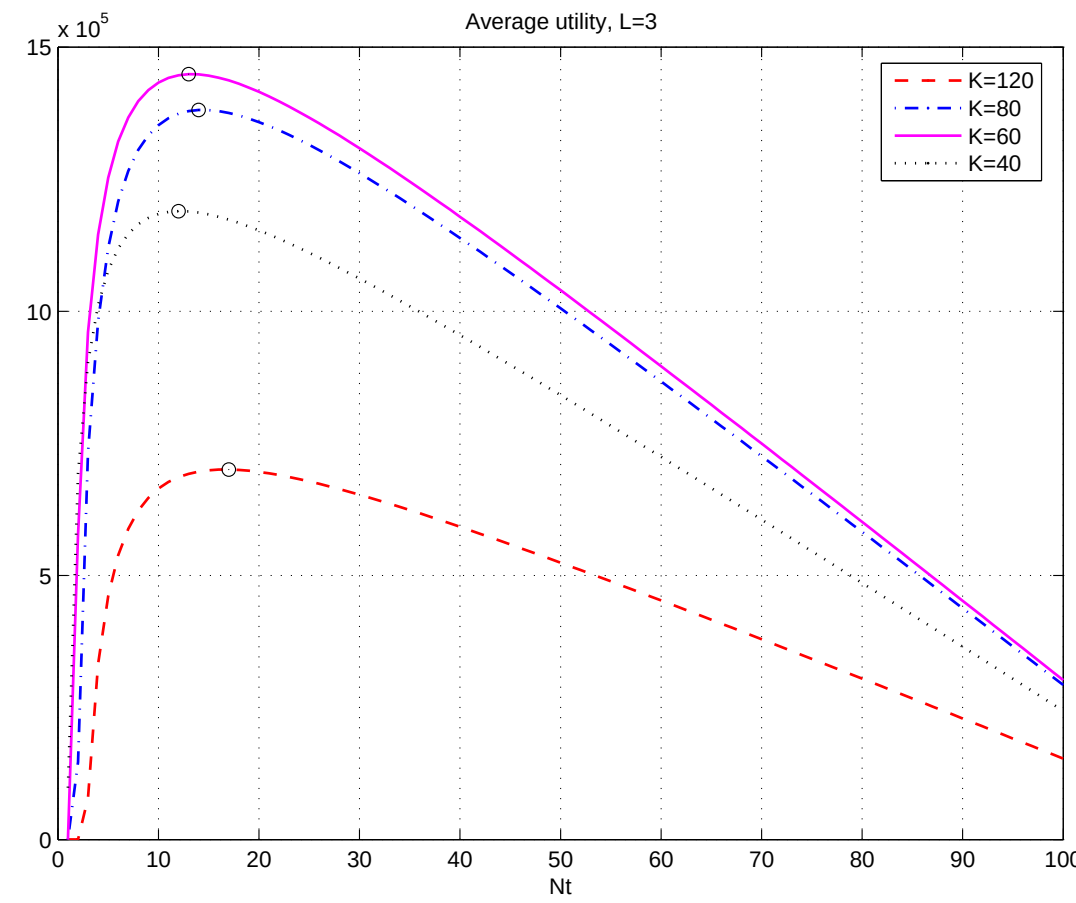

Fig. 4. Average utility versus training length for different values of the number of users $K$.

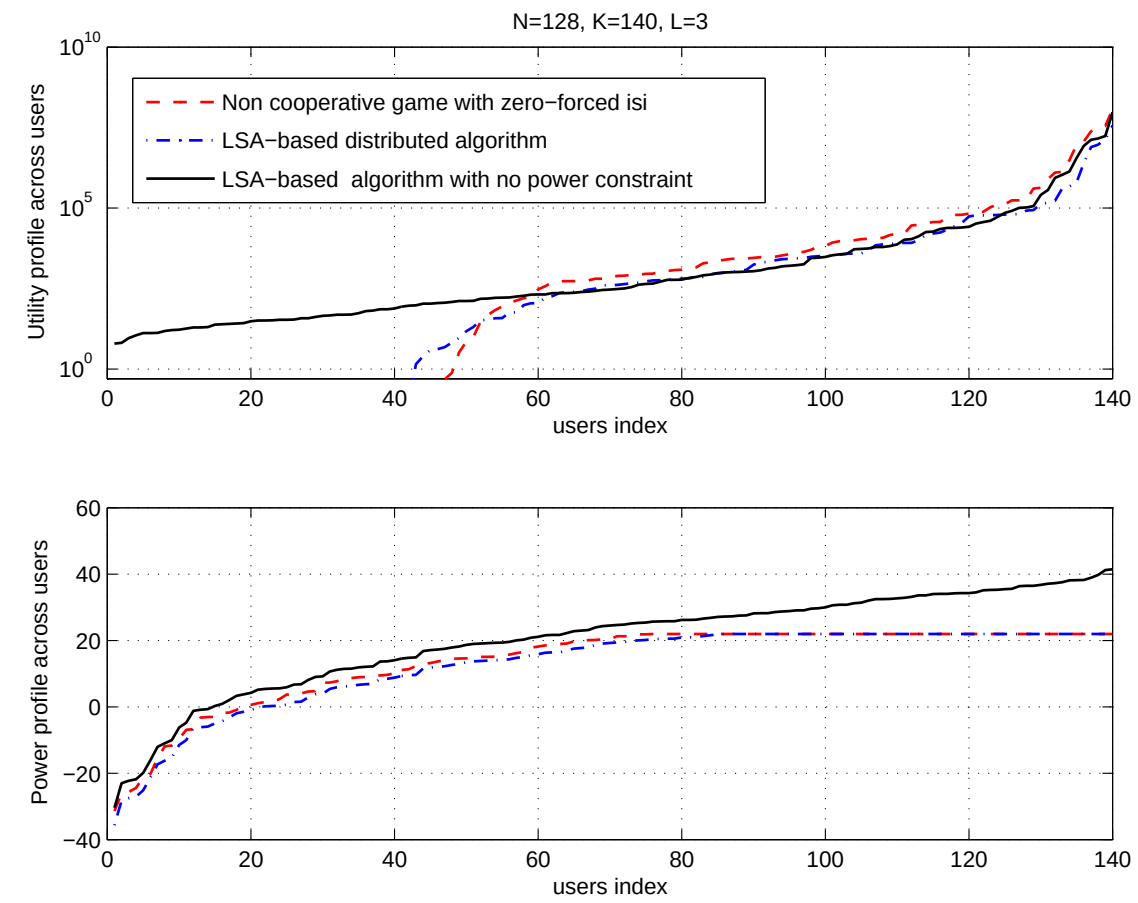

Fig. 5. Utility and power profile across users.

## VI. LSA-based distributed power control with constrained transmit power

Assume finally that there is a maximum allowed transmit power $P_{\max}$. Given the power profile of Eq. (11) we can count the number of users $u_2$ with transmit power larger than $P_{\max}$. It is natural to assume that the users transmitting at $P_{\max}$ will be the ones with the smallest channel gains, which, due to the lemma of Section III, can be approximated by $F^{-1}(\frac{K-l}{K})$, with $l = K-u_2+1, \ldots, K$. As a consequence, the generic $k$-th user will be affected by $u_1 = K-u_2$ users that are received with power $P_R$, and by $u_2$ users that are received with power $P_{\max}F^{-1}(\frac{K-l}{K})$, with $l = K-u_2+1, \ldots, K$. Denoting by $P_k$ the received power for the $k$-th user, the asymptotic $k$-th user SINR can be now written as

$$\gamma_k = \frac{P_k}{\frac{\mathcal{N}_0}{2} + \frac{u_1}{N}\frac{P_kP_R}{P_k+P_R\gamma_k} + \frac{1}{N}\sum_{i=K-u_2+1}^{K}\frac{P_kP_{\max}F^{-1}(\frac{K-l}{K})}{P_k+P_{\max}F^{-1}(\frac{K-l}{K})\gamma_k}}. \qquad (20)$$

Now, assuming for the moment that user $k$ is able to achieve its target SINR, i.e that $P_k = P_R$, the approximation $\frac{P_kP_R}{P_k+P_R\gamma_k} \approx \frac{P_k}{1+\gamma_k}$ , can be used in (20), and equating it to the target SINR $\bar{\gamma}$, we have

$$\frac{P_k}{\frac{\mathcal{N}_0}{2} + \frac{u_1}{N}\frac{P_k}{1+\bar{\gamma}} + \frac{1}{N}\sum_{i=K-u_2+1}^{K}\frac{P_kP_{\max}F^{-1}(\frac{K-l}{K})}{P_k+P_{\max}F^{-1}(\frac{K-l}{K})\bar{\gamma}}} = \bar{\gamma}. \qquad (21)$$

The above relation can now be solved numerically in order to determine the receive power $P_k$ for the $k$-th user; the actual transmit power for the $k$-th user is finally set according the rule $p_k = \min\left\{\frac{P_k}{\sum_{l=0}^{L-1}|\alpha_{k,l}|^2}, P_{\max}\right\}$. Note that again this algorithm requires knowledge of the channel gains only for the user of interest. Results in Fig. 5 show that the LSA-based power control algorithm tightly approximates the power profile determined by direct application of the proposition in Section II.

## VII. Conclusions

The problem of LSA-based power control for energy efficiency in asynchronous multipath CDMA channels has been considered in this paper. It has been shown that the use of LSA leads to simple and decentralized power control procedures, and to simple formulas predicting the performance profile and the operating point of a large wireless CDMA-based network. Numerical results have shown that the LSA-based power control policy approximates with good accuracy, even in not-so-large systems, the performance of conventional power-control algorithms requiring a much larger amount of prior information.